\documentclass[prc,showpacs,nofootinbib,preprint]{revtex4}
\usepackage[colorlinks]{hyperref}
\usepackage{epsfig,graphics,color}
\usepackage{subfigure}
\usepackage{amssymb}
\usepackage{amsmath}

\def\w2{\tilde w^2}
\def\ws2{1}

\newlength{\lslash}

\begin{document}
\title{On the consistency of recent QCD lattice data \\ of the baryon ground-state masses}
\author{M.F.M. Lutz and A. Semke}
\affiliation{GSI Helmholtzzentrum f\"ur Schwerionenforschung GmbH,\\
Planck Str. 1, 64291 Darmstadt, Germany}
\date{\today}
\begin{abstract}
In our recent analysis of lattice data of the
BMW, LHPC and PACS-CS groups we determined a parameter set of the
chiral Lagrangian that allows a simultaneous description of the baryon
octet and decuplet masses as measured by those lattice groups. The results on the baryon spectrum of
the HSC group were recovered accurately without their inclusion into
our 6 parameter fit. We show that the same parameter set provides an accurate
reproduction of the recent results of the QCDSF-UKQCD group probing the baryon masses at quite different quark masses.
 This shows a remarkable consistency amongst the different lattice simulations.
 With even more accurate lattice data in the near future
it will become feasible to determine all low-energy parameters relevant at N$^3$LO.
\end{abstract}

\pacs{25.20.Dc,24.10.Jv,21.65.+f}
 \keywords{Chiral extrapolation, Large-$N_c$, chiral symmetry, flavor $SU(3)$, Lattice QCD}
\maketitle

\section{Introduction}

The remarkable advance of QCD lattice simulations allows for the accurate determination
of low-energy constants characterizing effective Lagrangians that are formulated with hadronic degrees
of freedom and tuned to describe the low-energy reactions in QCD \cite{Procura:2006bj,Jenkins:2009wv,MartinCamalich:2010zz,PhysRevD.81.014503,Semke2012,Semke:2012gs,Bruns:2012eh,Shanahan:2012}.
An important example is the chiral Lagrangian with baryon octet and decuplet fields used to describe the meson-baryon scattering data \cite{Kaiser:1995eg,Becher1999,Lutz2002a,Mai:2009ce,Bruns:2010sv,Ikeda:2012au}. The latter is difficult at present to be described by lattice simulations directly.

Systematic applications of the chiral Lagrangian with three light flavors are hampered by the slow convergence
of a strict chiral expansion (see e.g.  \cite{LHPC2008,PACS-CS2008,MartinCamalich:2010fp,WalkerLoud:2011ab}). An acceleration of the 
convergence is often achieved by systematic summations that improve the realization of some selected basic physics principles. 
For instance, the use of the relativistic variants of the chiral Lagrangian
recovers naturally Lorentz covariance with its associated and specific analyticity properties of reaction amplitudes and hadronic
self energies. Also, the precise realization of coupled-channel unitarity leads to a plethora of novel applications that appear to grasp
many nontrivial physics issues (see e.g. \cite{Lutz:2003fm,Lutz:2005ip,Gasparyan:2010xz,Bruns:2010sv,Danilkin:2011fz}).

Recently it has been pointed out by the authors that in a computation of the baryon self energies it is crucial to insist
on their correct analytic structure \cite{Semke2005,Semke2007,Semke2012}. The emission of a virtual meson requires the baryon 
self energy to have branch cuts at the sums of meson and baryon masses. While the description of the strength of the branch cut requires dynamical information, the position of the branch point is model independently determined. The realization of such constraints is readily implemented by the use of
physical (or pole) masses in the evaluation of any loop diagram contributing to the baryon self energies.
While the use of physical masses is widespread in many applications of chiral Lagrangians, it was not been used before in the evaluation of the baryon
self energy. The reason may be some technical complications implied by this analyticity constraint. Indeed, given such a scheme the baryon masses
are to be determined numerically as a solution of coupled and non-linear sets of equations. For more details we refer to
our previous works  \cite{Semke2005,Semke2007,Semke2012}.

There is a further challenge in the accurate determination of the baryon masses as function of different light quark masses within an effective Lagrangian framework.
At the order where
one expects an accurate reproduction of the baryon massesб there are many a priori unknown low-energy constants that
surely cannot be determined from current lattice data. To overcome this problem the authors studied the consequences of large-$N_c$ QCD that were shown
to reduce the number of parameters significantly \cite{LutzSemke2010,Semke2012}. In a first explorative scenario we considered 14 parameters, where the class of the chiral symmetry preserving counter terms was switched off. In contrast to the conventional spirit
of lattice groups, we do not aim to predict the experimental baryon masses. Rather, eight out of the 14 parameters were used to recover the empirical baryon masses exactly. That leaves six free parameters to be adjusted to various lattice data. With those 6 parameters we successfully reproduced 73 lattice points from the BMW, LHPC, PACS-CS and HSC groups \cite{BMW2008,LHPC2008,PACS-CS2008,HSC2008}.

The purpose of this short letter is to document the implications of our parameter sets for the more recent results
from the QCDSF-UKQCD group \cite{Bietenholz:2011qq}.

\section{Chiral extrapolation of baryon masses}

In our previous work we performed three different fits \cite{Semke:2012gs}. The first one
considered the pion-mass dependence of the nucleon and omega masses of the BMW group together with the data of the LHPC group \cite{Durr:2011mp,WalkerLoud:2011ab}. The latter provided a high statistical study on various baryon mass splittings.
The second fit combined the PACS-CS and BMW data for the baryon masses. Finally, in the third fit a global reproduction of the above lattice data was achieved.
The parameters of those three fits are detailed in \cite{Semke2012,Semke:2012gs}. As the first non-trivial implication we observed that Set 2 and Set 3 reproduce the data
of the HSC group on the baryon octet and decuplet masses with a chi-square per point close to one, even though this data was not considered
in any of our fits.

We turn to the recent results of the QCDSF-UKQCD group. Results for all baryon octet and decuplet masses at various pion and kaon masses were
published in \cite{Bietenholz:2011qq}. Following our previous strategy we determine the light quark masses from a chiral expansion of the meson masses at next-to-leading order \cite{Semke2012}. In the following we consider the simulation data that rely on their largest lattice only,
i.e. the $32^3 \times 64$ case. Here we do not correct for possible, but small, lattice size effects. For the lattice scale $a$
the estimate
\begin{eqnarray}
a = ( 0.0765 \pm 0.0015 )\,{\rm fm}
\label{lattice-scale}
\end{eqnarray}
was given in \cite{Bietenholz:2011qq}.

\begin{figure}[t]
\centering
\includegraphics[width=8cm,clip=true]{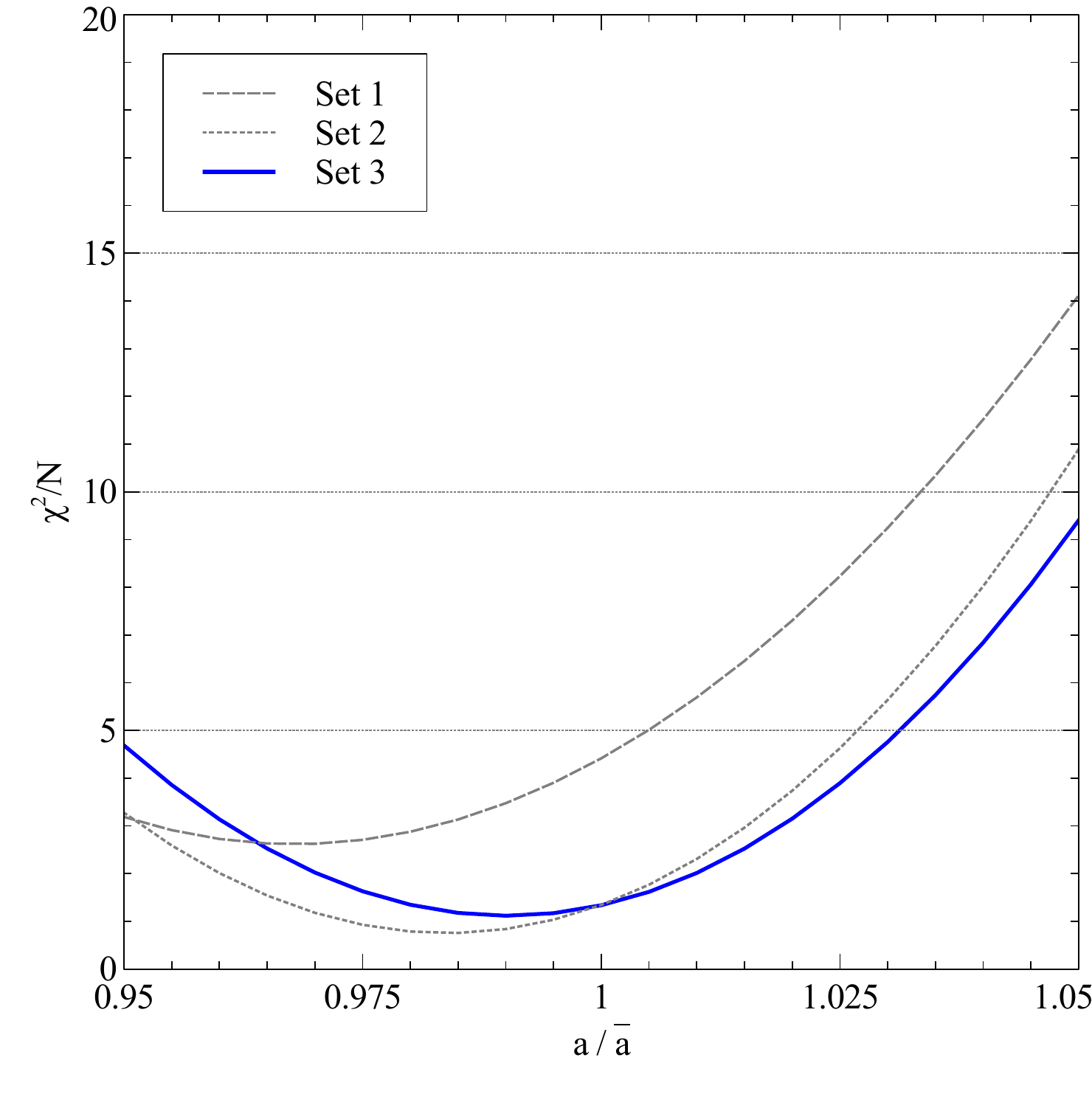}
\includegraphics[width=8cm,clip=true]{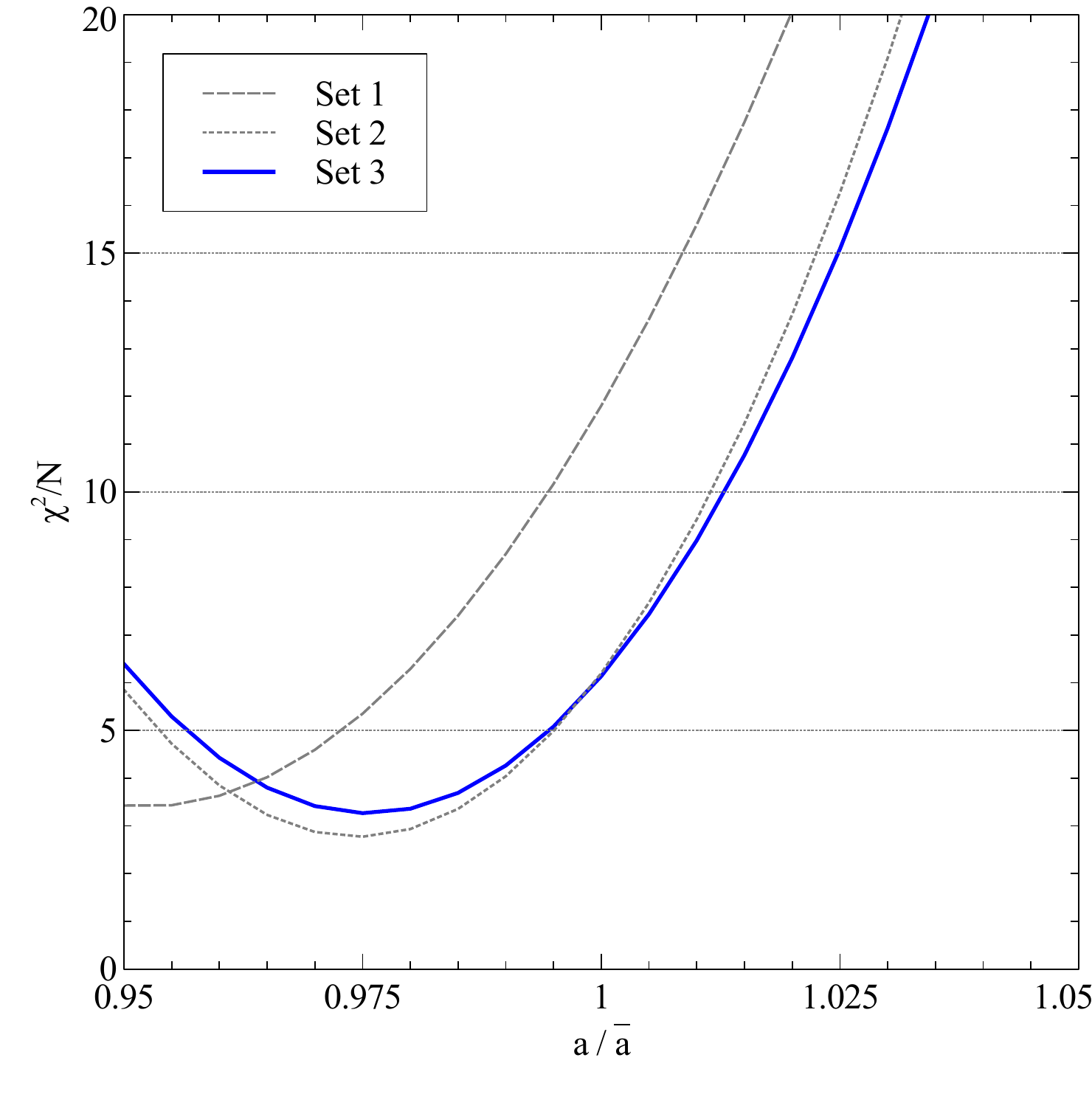}
\caption{$\chi^2/N$ as a function of the lattice spacing $a/\bar a$ with $\bar a=0.0765$ fm. 
Lattice data is taken from \cite{Bietenholz:2011qq}. The left and right-hand panels correspond to 
the flavor symmetric and asymmetric cases, respectively.}
\label{fig:1}
\end{figure}

The QCDSF-UKQCD data is grouped into two distinct classes. There are three data sets in the flavor SU(3) limit and further three data sets with flavor breaking, but keeping the singlet combination of the quark masses constant.
In the initial step we used our previously determined parameters sets ( Set 1 - 3 ) and compute the
implied $\chi^2/N$ of the two classes of data as a function of the lattice scale. Our chi square is based on statistical errors only and therefore  should be viewed as an upper bound for a full chi square function. We consider a variation of $5\%$ around the
central value (\ref{lattice-scale}). In Fig. \ref{fig:1} the two chi square functions are shown with respect to our three sets of parameters.
We find remarkable that the chi square functions for the flavor symmetric data (l.h.p.) computed with Set 2 and Set 3 show their minima at a scale that is about (1.0-2.5) $\%$ smaller than the central value given in (\ref{lattice-scale}). This is within the estimated range determined by the QCDSF-UKQCD group. The flavor symmetric QCDSF-UKQCD data clearly favors Set 2 and Set 3 over Set 1, which has a significantly larger
chi square. The minimal $\chi^2/N$ for Set 2 and Set 3 is close to one. For the flavor breaking data  set (r.h.p.) the minimal $\chi^2/N$ of all three fits is about 3, leaving room for further improvements. Still Set 1 is disfavored since the optimal lattice scale appears outside the estimated range (\ref{lattice-scale}). We find interesting that the flavor asymmetric data sets lead to a much steeper chi square function as compared to the flavor symmetric data. This may be a useful observation for future high precision determinations of the lattice scale.

\begin{figure}[t]
\centering
\includegraphics[width=8cm,clip=true]{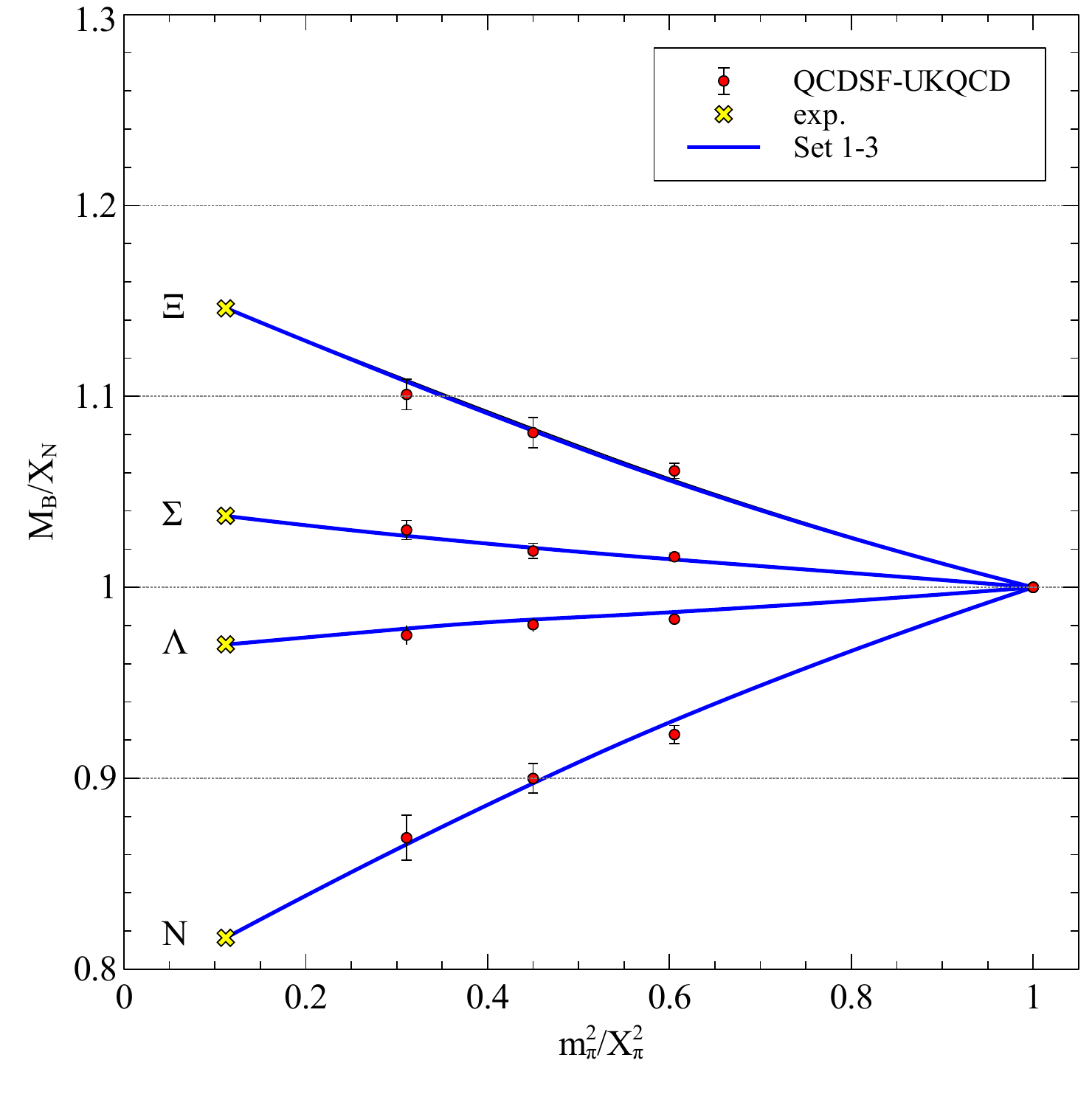}
\includegraphics[width=8cm,clip=true]{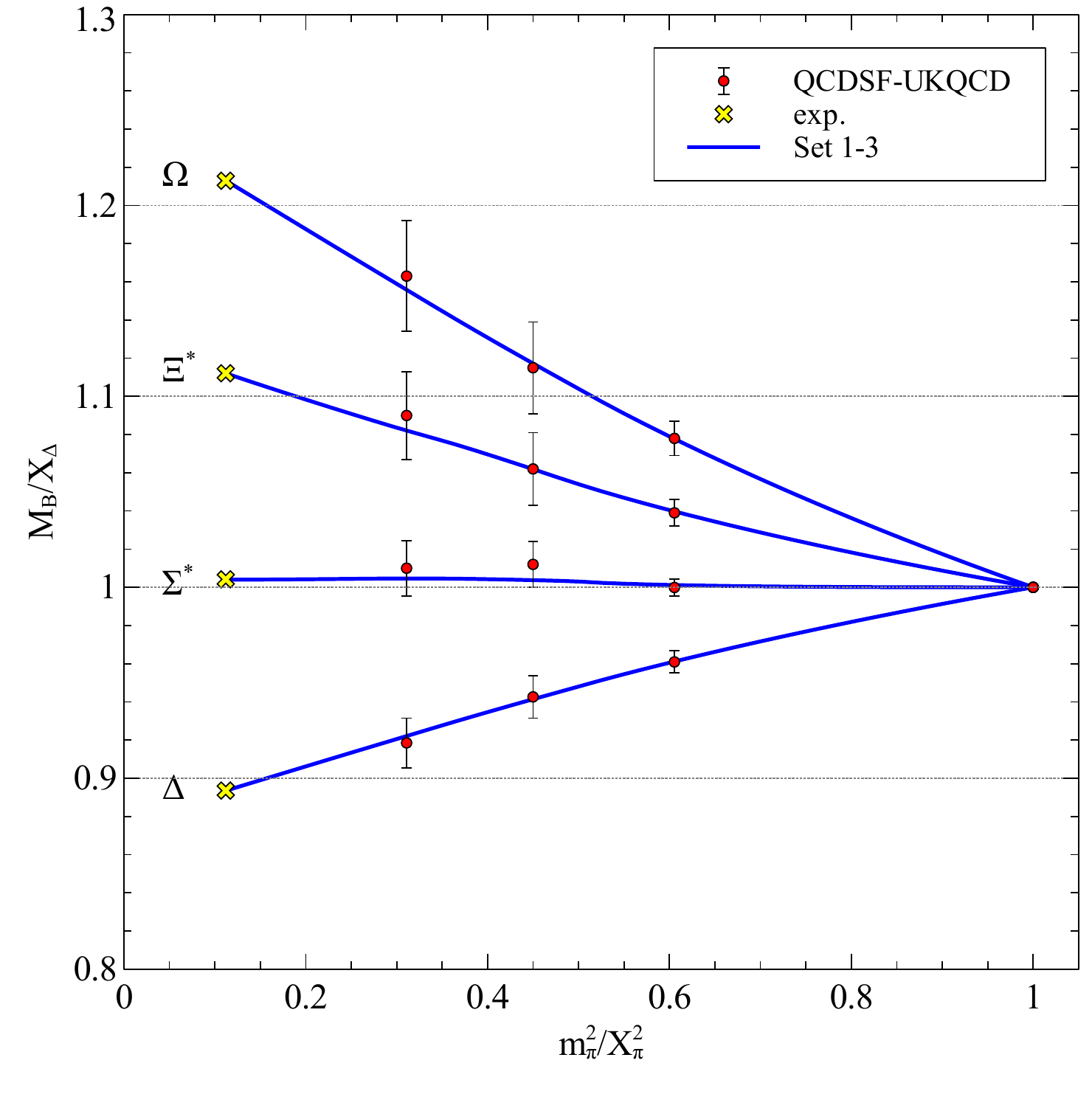}
\caption{Chiral extrapolation of ratios of baryon masses as explained in the text. The QCDSF-UKQCD data  \cite{Bietenholz:2011qq} were 
not considered in the determination of the three parameters sets. }
\label{fig:2}
\end{figure}

The QCSDF-UKQCD group suggests to consider specific ratios of the baryon masses that are expected to allow for a 'simpler' chiral extrapolation.
Ratios of the baryon masses over $X_N$ or $X_\Delta$ are considered as a function of $m^2_\pi/X_\pi^2$ with
\begin{eqnarray}
X_\pi^2 = \frac{1}{3}\,\Big( 2\,m_K^2+m_\pi^2 \Big) \,, \qquad X_N = \frac{1}{3}\,\Big( M_N + M_\Sigma + M_\Xi \Big)
\,, \qquad X_\Delta = \frac{1}{3}\,\Big( 2\,M_\Delta + M_\Omega  \Big)\,,
\label{def-X}
\end{eqnarray}
where isospin averaged masses are assumed. The merit of such ratios is twofold. Firstly, they are independent of the lattice scale. Secondly, they were shown to have an almost linear dependence on the variable $m_\pi^2/X_\pi^2$. In Fig. \ref{fig:2} the lattice results for the octet and decuplet ratios data are recalled.

Given the errors for the various ratios, we computed $\chi^2/N$ with respect to the our parameter scenarios. For all three cases we obtain values below one. The continuous blue lines in Fig. \ref{fig:2} follow from a computation where the sum of the three light quark masses is kept constant at its physical value. The three parameter scenarios lead to indistinguishable predictions in Fig. \ref{fig:2}. The lines are remarkably close to the QCDSF-UKQCD data points. Similar results, however, only for the octet ratios, were recently obtained in \cite{Shanahan:2012,Bruns:2012eh}.

\section{Conclusions}

A controlled application of the chiral Lagrangian with baryon octet and decuplet fields was presented.
In contrast to the conventional strategy, our aim is not to predict the empirical baryon masses from
the lattice data, rather we use the empirical baryon masses as additional constraints to learn more about the various
low-energy parameters of the chiral Lagrangian. The baryon self energies were used at N$^3$LO, where we insisted on the
use of the physical masses in the loop functions. The number of parameters was reduced in application of sum rules that follow from
QCD in the limit of a large number of colors. In our current setup we are left with 6 unknown parameters only.
In our previous works we established 3 parameter scenarios based on a simultaneous analysis of data from the BMW, LHPC and PACS-CS groups.
Given those parameter sets we computed the quark-mass dependence of the baryon masses as measured recently by the QCDSF-UKQCD group.
We find a remarkable agreement between our prediction based on the chiral Lagrangian and the QCD lattice simulations of the QCDSF-UKQCD group.
We arrived at a 6 parameter reproduction of alltogether 105 data points from 5 different QCD lattice groups.

\bibliography{literatur}
\end{document}